\documentclass[%
 aip,
 jmp,%
 amsmath,amssymb,
reprint,%
 fleqn,%
groupedaddress]{revtex4-1}

\usepackage{graphicx}
\usepackage{dcolumn}
\usepackage{bm}
\begin{document}
\hfill BU-HEPP-18-05
\title{Multiquark States in the Thomas-Fermi Quark Model and on the Lattice}

\author{Walter Wilcox$^*$ and Suman Baral$^\dagger$}

\address{Physics Department, Baylor University,\\
Waco, TX 76798-7316 USA\\
$^*$E-mail: walter\_wilcox@baylor.edu\\
$^\dagger$E-mail: suman\_baral@baylor.edu, suman@neuralinnovations.io}

\begin{abstract}
We describe work being done at Baylor University investigating the possibility of new states of mesonic matter containing two or more quark-antiquark pairs. To put things in context, we begin by describing the lattice approach to hadronic physics. We point out there is a need for a quark model which can give an overall view of the quark interaction landscape. A new application of the Thomas-Fermi (TF) statistical quark model is described, similar to a previous application to baryons. The main usefulness of this model will be to detect systematic energy trends in the composition of the various particles. It could be a key to identifying {\it families} of bound states, rather than individual cases. Numerical results based upon a set of parameters derived from a phenomenological model of tetraquarks are given.
\end{abstract}


\maketitle

\section{Introduction and Motivation}\label{intro}

\begin{figure}
\begin{center}
\includegraphics[width=3.5in]{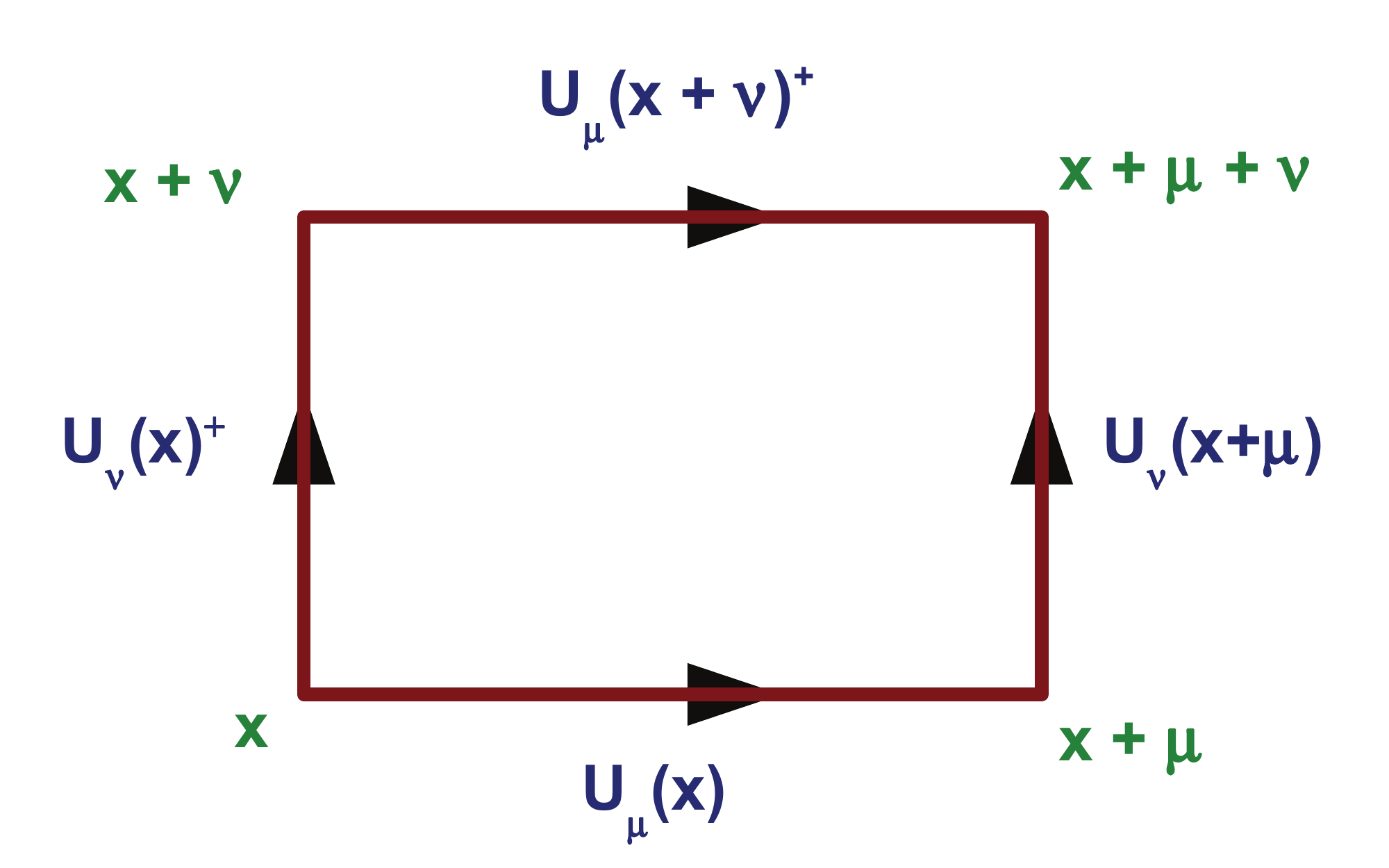}
\caption{The link variables on a hypercubic lattice.}
\label{fig2}
\end{center}
\end{figure}

\begin{figure}
\begin{center}
\includegraphics[width=3.0in]{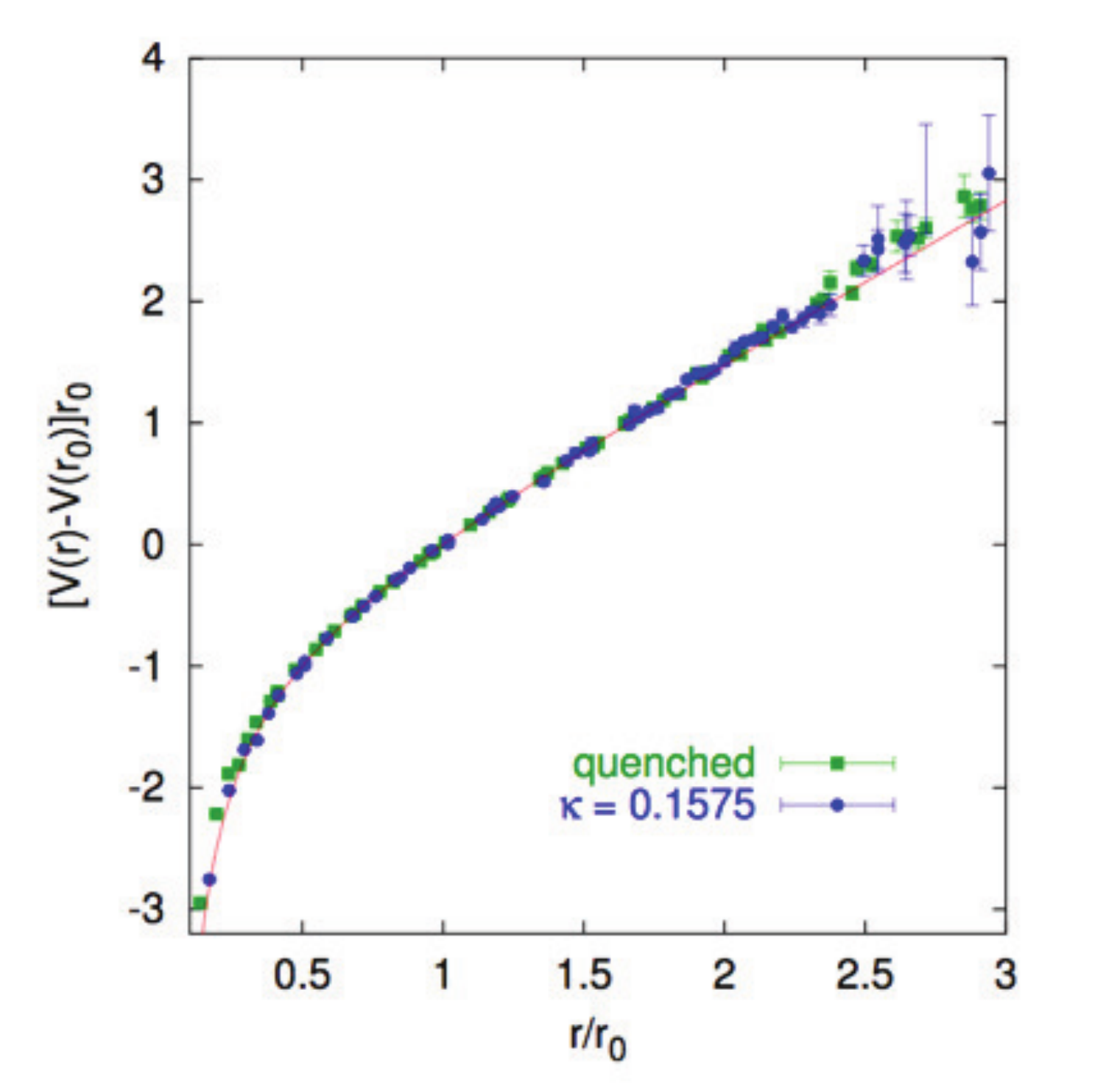}
\caption{The heavy quark-antiquark potential for quenched and dynamical configurations.}
\label{fig3}
\end{center}
\end{figure}

\begin{figure}
\begin{center}
\includegraphics[width=1.90in]{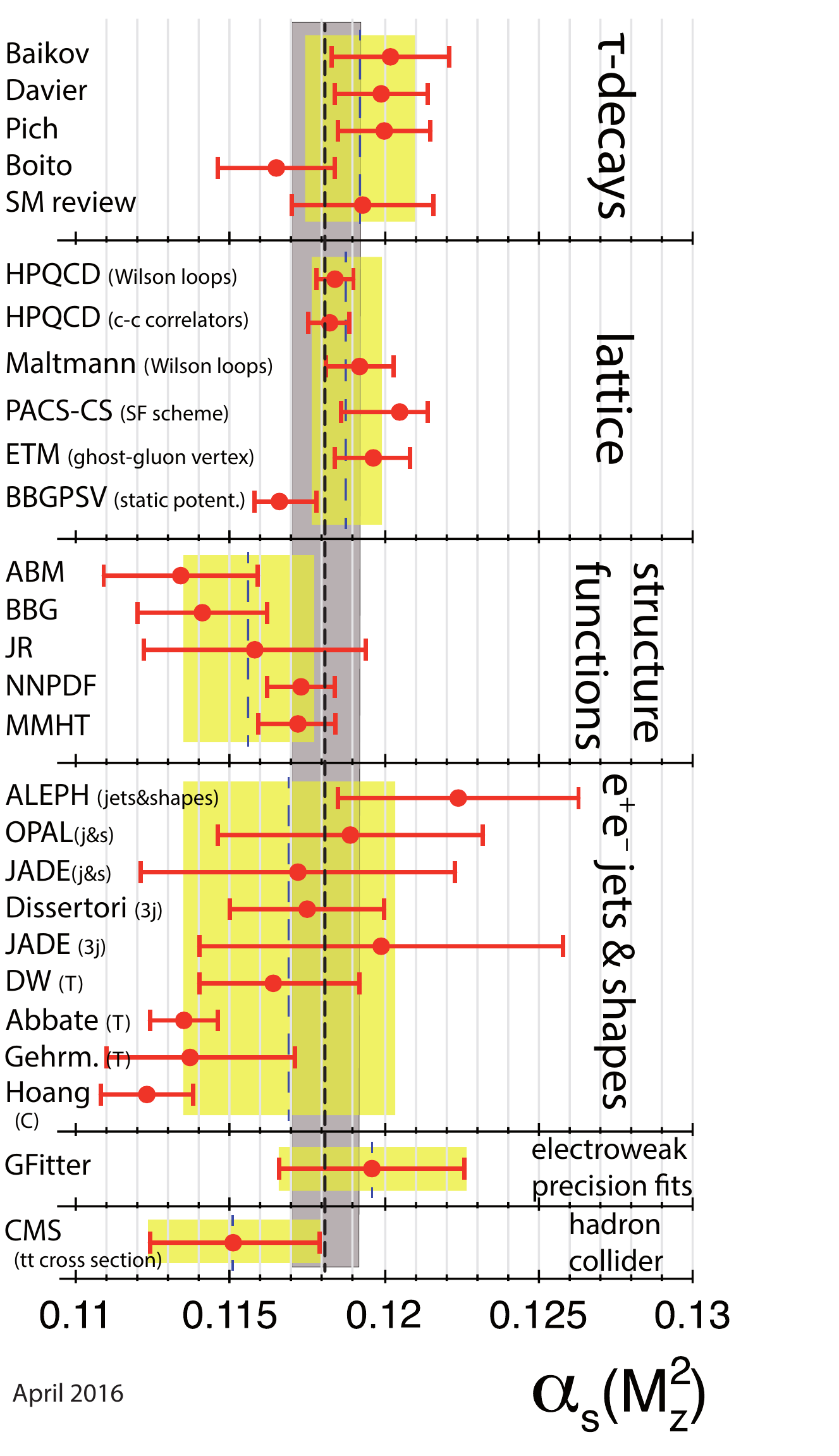}
\caption{Recent measurements of the strong coupling constant, $\alpha_s$.}
\label{fig4}
\end{center}
\end{figure}

\begin{figure}
\begin{center}
\includegraphics[width=4.5in]{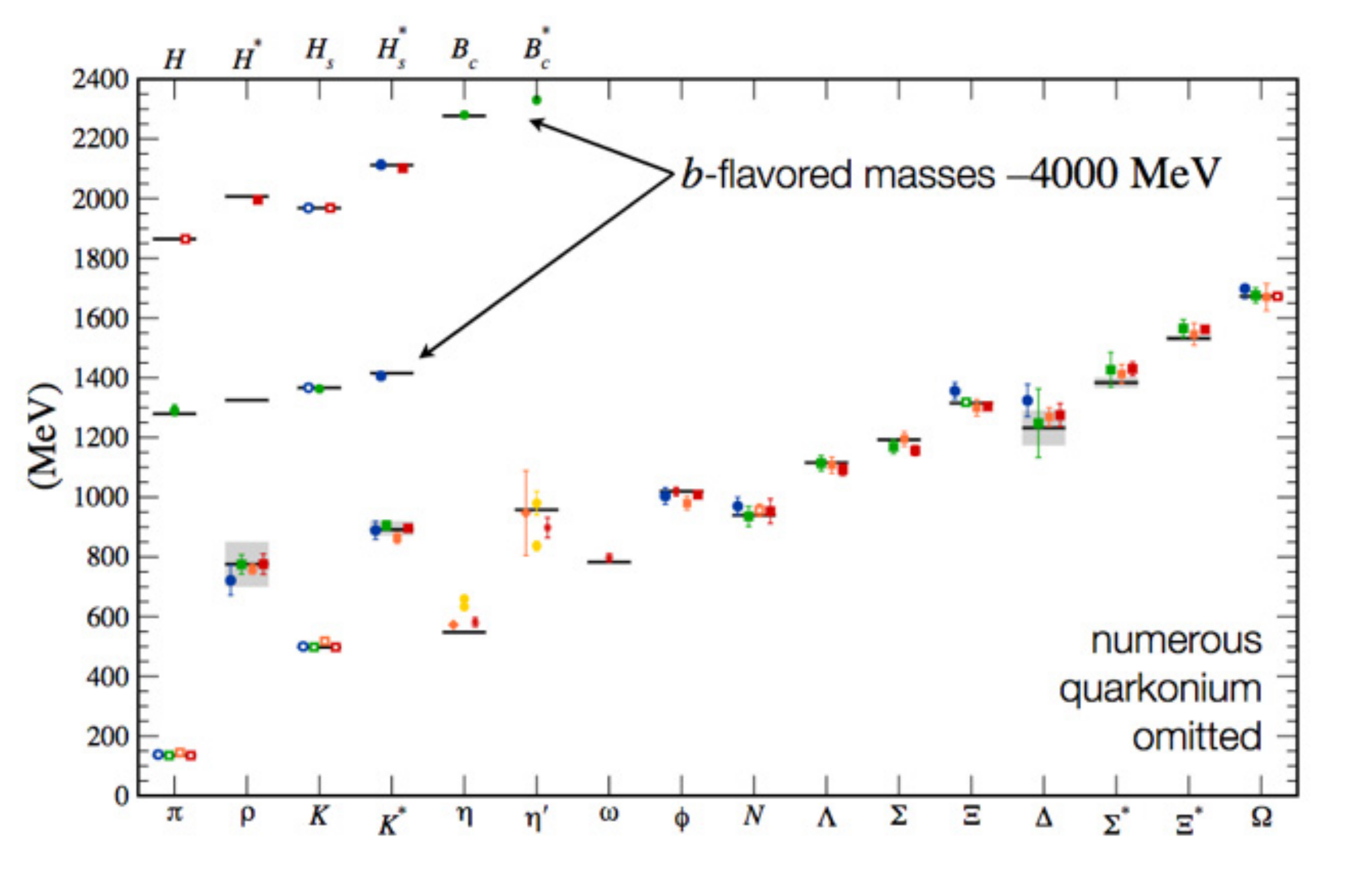}
\caption{A compilation of lattice QCD hadron mass calculations.}
\label{fig5}
\end{center}
\end{figure}

Lattice Chromodynamics (QCD) is the main tool used by particle physicists to investigate the properties of baryons and mesons within the context of the strong interactions. The technology and algorithms of lattice applications are constantly improving. The path integral approach, pioneered by Feynman, is done automatically via Monte Carlo simulation. The quark degrees of freedom, including color, spin and particle/antiparticle, are incorporated into a quark \lq\lq mass matrix", which is used to define quark and hadron propagation functions. The lattice scale is set by observed renormalization group behavior. In addition, the lattice \lq\lq link" variables as depicted in Fig.~\ref{fig2} play the role of the gluon degrees of freedom. Fig.~\ref{fig3} shows the heavy quark-antiquark potential for mesons extracted from lattice gluonic combinations\cite{heavy}. Lattice configurations can be quenched or dynamical. Quenched lattice configurations suppress background quark-antiquark loops in order to limit computer time requirements. Dynamical or nonquenched calculations can accommodate light, strange and charmed quark loops, and are now used in all realistic lattice calculations. Fig.~\ref{fig4} gives the present Particle Data Group (PDG) summary of extractions of the strong coupling constant, showing that lattice QCD now results the smallest error bars\cite{PDG1}. Finally, Fig.~\ref{fig5} gives a contemporary depiction of the state of lattice QCD spectrum calculations, showing light quark baryons and mesons as well as states with both hidden and explicit charm and bottom\cite{Kron}. These results are impressive, and confirm that QCD is the correct theory of the strong interactions. 


Tetraquark and pentaquark states are now known to exist from Belle\cite{Belle0,Belle,B2,B3}, BESIII\cite{BESIII,BES2}, LHCb\cite{LHCb00,LHCb,LHCb1,LHCb0,LHC2}, and other collaborations. Are there also states of many more quarks? Could there be an analog to nuclear/atomic systems for heavy/light quark systems? As the quark content increases, it becomes computationally expensive and time-intensive to do the lattice calculations. Every state must be investigated separately, which means a great deal of analysis on Wick contractions and specialized computer coding in lattice QCD. In addition as one adds more quarks the states will become larger and the lattice used must also increase in volume.

There is a need for quark models which can help lead expensive and spatial-limited lattice QCD calculations in the right direction in the search for high quark states. The Thomas-Fermi (TF) statistical model has been amazingly successful in the explanation of atomic spectra and structure, as well as nuclear applications. The atomic applications by Schwinger and Englert have brought it to its highest point. Our group has adopted the TF model and applied it to collections of many quarks\cite{WW1,WW2,Baral}. One would expect that the TF quark model would become increasingly accurate as the number of constituents is increased, as a statistical treatment is more justified. The main usefulness will be to detect systematic trends as the parameters of the model are varied. It could also be key to identifying {\it families} of bound states, rather than individual cases. We have now extended the TF quark model to mesonic states in order to investigate the stability of families built from some existing mesons and observed new exotic states, concentrating on heavy-light quark combinations. Although our model is nonrelativistic, we will see that this assumption is actually numerically consistent as quark content is increased.

\section{TF Quark Model}\label{model}

The Thomas-Fermi model treats particles as a Fermi gas at $T=0$. It builds in Fermi statistics, but is not fully quantum mechanical. It does not have a quantum mechanical wave function, but instead a central function related to particle density which is determined by filling states up to the Fermi surface at each physical location. It gives accurate atomic binding energies for large numbers of electrons in atomic systems. It is exactly what is called for in this situation with many quarks. The TF model is better than \lq\lq bag models"\cite{Mit,Mads1,Mads2}, which do not intrinsically include the Coulomb interactions for large numbers of particles. The challenge we face of course is that we are extrapolating from small numbers of particles, where the model is less accurate, to large numbers. However, our goal is to detect systematic trends in particle states, which we believe will be manifest in this many particle theory.

Explicit spin interactions can be included in the model, but some limitations are present. In the atomic TF model (and in nuclear applications) the up and down spin 1/2 states are treated as degenerate. One cannot do this in particle physics! In our treatment, the spin quantum number is separated out as another \lq\lq flavor". Spin of course is not conserved in particle interactions, only the total angular momentum corresponds to a conserved quantum number. However, this breaks rotational symmetry. It is best to keep the states in a maximal spin \lq\lq up" or \lq\lq down" orientation. Then classical and quantum states are \lq\lq maximally compatible". Such a treatment has been developed for baryons.\cite{WW2} In our present limited treatment for mesons, we have not yet included explicit spin interactions, but we can take one level of spin degeneracy into account.

The types of interactions between the particles can be categorized as\cite{Sikivie}:\\

\noindent
\textbf{Color-Color Repulsion (CCR)}
Interactions between quarks with same colors is repulsive with coupling constant $4/3g^2$. The interactions are red-red ($rr$), green-green ($gg$) and blue-blue ($bb$).\\

\noindent
\textbf{Color-Color Attraction (CCA)}
Interactions between quarks with different colors is attractive with coupling constant $-2/3g^2$. The interactions are $rb$, $rg$, $bg$, $br$, $gr$,  and  $gb$.\\

\noindent
\textbf{Color-Anticolor Repulsion (CAR)}
Interactions between quarks and antiquarks with different color/anticolors is repulsive with coupling constant $2/3g^2$. The interactions are $r\bar{b}$, $r\bar{g}$, $b\bar{g}$, $b\bar{r}$, $g\bar{r}$, and $g\bar{b}$.\\

\noindent
\textbf{Color-Anticolor Attraction (CAA)}
Interactions between quarks and antiquarks with same color/anticolors is attractive with coupling constant $-4/3g^2$. The interactions are $r\bar{r}$, $b\bar{b}$ and $g\bar{g}$.\\

\noindent
\textbf{Anticolor-Anticolor Repulsion (AAR)}
Interactions between antiquarks with same anticolors is repulsive with coupling constant $4/3g^2$. The interactions are $\bar{r}\bar{r}$, $\bar{b}\bar{b}$  and  $\bar{g}\bar{g}$.\\

\noindent
\textbf{Anticolor-Anticolor Attraction (AAA)}
Interactions between antiquarks with different anticolors is attractive with coupling constant $-2/3g^2$. The interactions are $\bar{r}\bar{b}$, $\bar{r}\bar{g}$, $\bar{b}\bar{g}$, $\bar{b}\bar{r}$, $\bar{g}\bar{r}$, and $\bar{g}\bar{b}$.\\

Table~\ref{Table1} shows the various color-averaged couplings within mesons with $\eta$ number of quark-antiquark pairs.
\begin{table}
{\begin{tabular}{ |c|c|c|c|}
\hline
Interaction type & Symbol  & Coupling & Interaction probability\\ 
\hline
 CCR & $P_{ii}$ &  $ \frac{4}{3} g^2$ & $\displaystyle\frac{(\eta -1 )} { 18(2\eta -1)}$  \\
\hline 
 CCA & $P_{ij}, i\ne j$  &  $ -\frac{2}{3} g^2$ & $\displaystyle\frac{\eta -1 } { 9(2\eta -1)}$\\ 
 \hline
 CAR & $\bar{P}_{ij},i\ne j$ &  $ \frac{2}{3} g^2$& $\displaystyle\frac{2(\eta -1 )}  {9(2\eta -1)}$\\
\hline 
 CAA & $\bar{P}_{ii}$ & $ -\frac{4}{3} g^2$  &$\displaystyle\frac{(\eta +2 )}{  9(2\eta -1)}$\\ 
 \hline
 AAR & $\bar{\bar P}_{ii}$ &$ \frac{4}{3} g^2$ & $\displaystyle\frac{(\eta -1 )} {18(2\eta -1)}$\\
\hline 
 AAA & $\bar{\bar P}_{ij},i\ne j$ & $ -\frac{2}{3} g^2$ &$\displaystyle\frac{(\eta -1 )}  {9(2\eta -1)}$\\ 
\hline
\end{tabular}}
\caption{The coupling constants and probabilities for certain types of quark and antiquark interactions in mesons.}
\label{Table1}
\end{table}
We find that, on average, quarks only interact with antiquarks in such systems; i.e., the sum of the products of the couplings and probabilities for {\bf CCR}, {\bf CCA} as well as {\bf AAR}, {\bf AAA} interactions vanish. If we add the remaining product of coupling and probabilities from Table~\ref{Table1}, we get $-{\frac{4}{3}g^2}/{(2\eta -1)}$, very similar to the baryon case.\cite{WW2} The negative sign indicates that the system is attractive because of the collective residual color coupling alone, even in the absence of volume pressure. This gives rise to a type of matter that is bound, but does not correspond to confined mesonic matter, as discussed in Ref.~\citenum{WW1}. We are interested in confined matter and will need to add a bag vacuum pressure term to the energy to enforce this.\cite{Mit} The TF differential equation is constructed and systems with heavy-light quark content are examined. Three types of heavy quark/antiquark ($Q$ or $\bar{Q}$) light quark/antiquark ($q$ or $\bar{q}$) mesonic systems are defined and investigated. Charmonium ($Q\bar{Q}, Q\bar{Q}Q\bar{Q}$, etc), $Z$-meson ($\bar{Q}Q\bar{q}q, \bar{Q}Q\bar{q} q\bar{Q}Q\bar{q}q$, etc.) and $D$-meson ($\bar{Q}q, \bar{Q}q\bar{Q}q $, etc.) family types are examined. These will be referred to as \lq\lq Case 1", \lq\lq Case 2" and \lq\lq Case 3", respectively, in the following. We will examine the charmed quark case below.

The phenomenological parameters we need for our model are the strong coupling constant $\alpha_s$, the bag constant $B$, the charm quark mass, $m_c$, as well as the light quark mass, $m^1$. Previously, we used baryon phenomenology to obtain these parameters\cite{Baral}, but we now wish to attempt a more realistic fitting using mesonic states. Since we do not yet include spin interactions in our model, we need to weight spin-split states to \lq\lq remove\rq\rq this interaction for our model fits. 
Assuming the interactions are proportional to a spin splitting term, $\left< \vec{S}_1\cdot\vec{S}_2 \right>$ for two spin 1/2 particles, we weighted masses of $1S$ states such that 
\begin{equation*}
\frac{1}{4} \left(\eta_c\left(1S\right)\right) + \frac{3}{4}  \left(J/\Psi\left(1S\right)\right) = 3069 \text{ MeV},
\end{equation*}
for charmonium. For  the mass of  the $D$ meson we weight particles such that
\begin{equation*}
\frac{1}{4} \left(D\right) + \frac{3}{4}  \left(D^*\right) =  1973 \text{ MeV},
\end{equation*}
where we are also averaging over charge states of $D$ and $D^*$. In addition, for the Case 2 mass we spin-weight the $J=1$, $C=-1$ tetraquark states as\cite{PDGReview}
\begin{eqnarray*}
\frac{1}{4} \left(Z_c(3900)\right) + \frac{3}{4}  \left(X(4020)\right) =  3990 \text{ MeV}.
\end{eqnarray*}
This weighting comes from a model where the light quark spin dominates the mass splitting of these two spin 1 states. We solved the differential equations using an iterative implementation of {\bf NDSolve} in {\it Mathematica}. We searched parameter space such that the model $\chi^2$ was minimized using a grid search, obtaining $\sqrt{\chi^2}= 1.05$ MeV in the mass evaluations, an almost perfect fit. We obtained $\alpha_s = 0.217$, $B^{1/4} =103.5$ MeV and charm quark mass $m_c=1530$ MeV. Our light quark mass, $m^1=306$ MeV, we take from our previous TF baryon spectrum fit.\cite{WW2} Note that our conference paper Ref.~\citenum{Baral} contains a numerical error in the calculation of the Case 3 type family binding energies, which is corrected here. A more detailed explanation of our parameter fitting assumptions will be given in Ref.~\citenum{future}. Note we have not yet completed a more comprehensive examination of the physics associated with inclusion of the $b$-quark sector.

We will examine TF spatial functions and energies for the three cases defined above. In nuclear physics, one inspects the binding energy per nucleon in order to assess the relative stability of a given nucleus. We will do a similar investigation here. Thus, the important figure of merit in these evaluations is the total energy per quark, for if this increases as one adds more pairs, the family is unstable under decay to lower family members, whereas if it decreases, the family is stable. The actual quark mass dependence does not play a role in these considerations and so will not be included in our energy plots. The dengeneracy status of the quarks will play an important role in these considerations. For $\eta$ pairs of quarks, we have
\begin{align*}
&\text{Case 1:}\quad n\cdot g=\eta,\\
&\text{Case 2:} \quad n_1\cdot g_1+n_2\cdot g_2=\eta,\\
&\text{Case 3:} \quad n_1\cdot g_1=\eta, \quad n_2\cdot g_2=\eta.
\end{align*}
$n, n_1, n_2$ are the number particles in a given state, and $g, g_1, g_2$ are degeneracies. The index \lq\lq 1" refers to light quarks and \lq\lq 2" refers to heavy quarks. For charmed quarks, $g_2=1,2$ only, whereas for light quarks we may have $g_1=1,2,3,4$.

\section{Model Results}

\begin{figure}
\begin{center}
\includegraphics[width=4.5in]{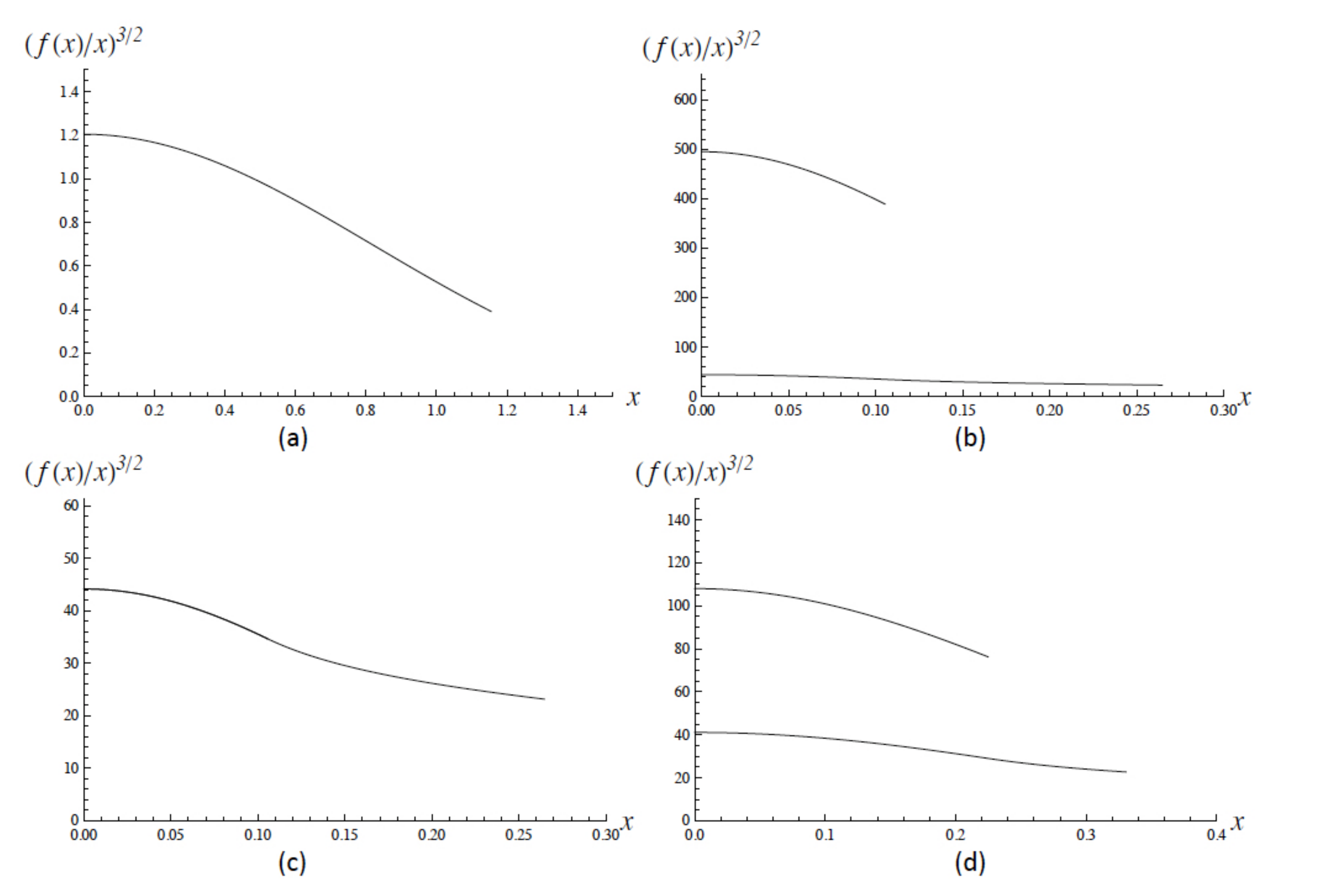}
\caption{TF density functions as a function of dimensionless distance $x$ for the various cases: (a) Charmonium (Case 1) (b) $Z$ mesons (Case 2) (c) Light quark density function for $Z$ mesons (d) $D$ meson (Case 3).}
\label{fig6}
\end{center}
\end{figure}

First, let us discuss the behavior of the particle density wave functions. We use the dimensionless parameter $x$ such that $r=Rx$ where
\begin{equation*}\label{fn2}
R\equiv \frac{a}{(8\alpha_{s}/3)}\left(\frac{3\pi\eta }{2}\right)^{2/3}.
\end{equation*}
$\alpha_s$ is the strong coupling constant and $a\equiv \hbar/(m^1 c)$. The particle density function of charmonium, proportional to $(f(x)/x)^{3/2}$, drops smoothly with increase in distance and has a discontinuity at the boundary due to the volume pressure, as seen in Fig.~\ref{fig6}(a). The density function of $Z$-mesons has a long tail for the light quarks, while for the heavy charmed quark the value is large and is concentrated near the origin, as seen in Fig.~\ref{fig6}(b). This suggests an atomic-like structure with heavy charm, anti-charm quarks at the center while light quarks and antiquarks spread out like electrons. Fig.~\ref{fig6}(c) is an enlargement of the density function of the light quark wave function for the $Z$-meson. It drops down abruptly until it reaches the boundary of the heavy quark wave function, then inflects and cuts off. In the case of $D$-mesons, Fig.~\ref{fig6}(d), the density function of light and heavy quarks are relatively closer. We increased the quark content and compared density functions of a family of multi-mesons in all three cases. We observed similar density functions for a given multi-meson family regardless of the quark content.

\begin{figure}
\begin{center}
\includegraphics[width=4.0in]{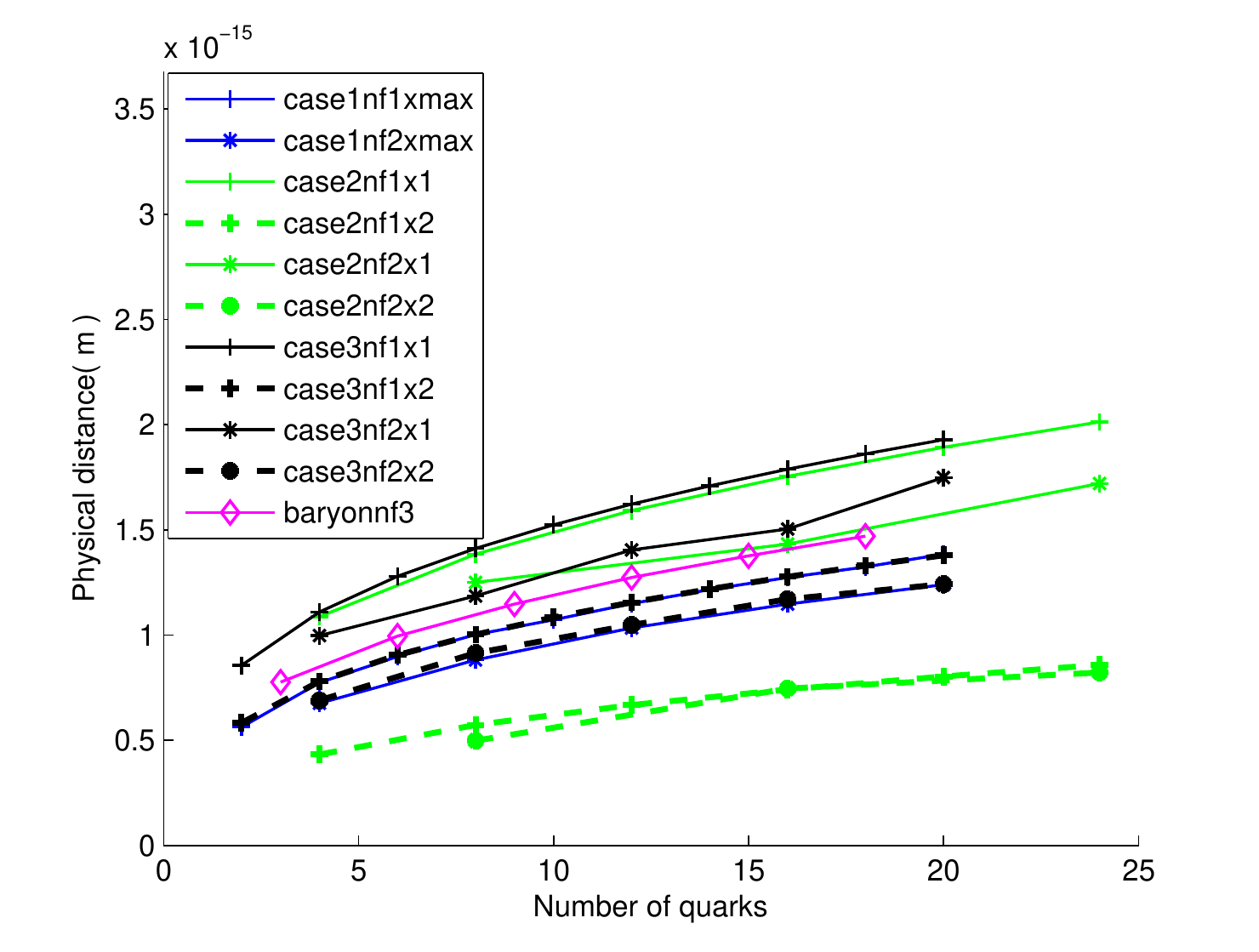}
\caption{Physical radius of the various mesons in this study versus quark content.}
\label{fig7}
\end{center}
\end{figure}

\begin{figure}
\begin{center}
\includegraphics[width=3.8in]{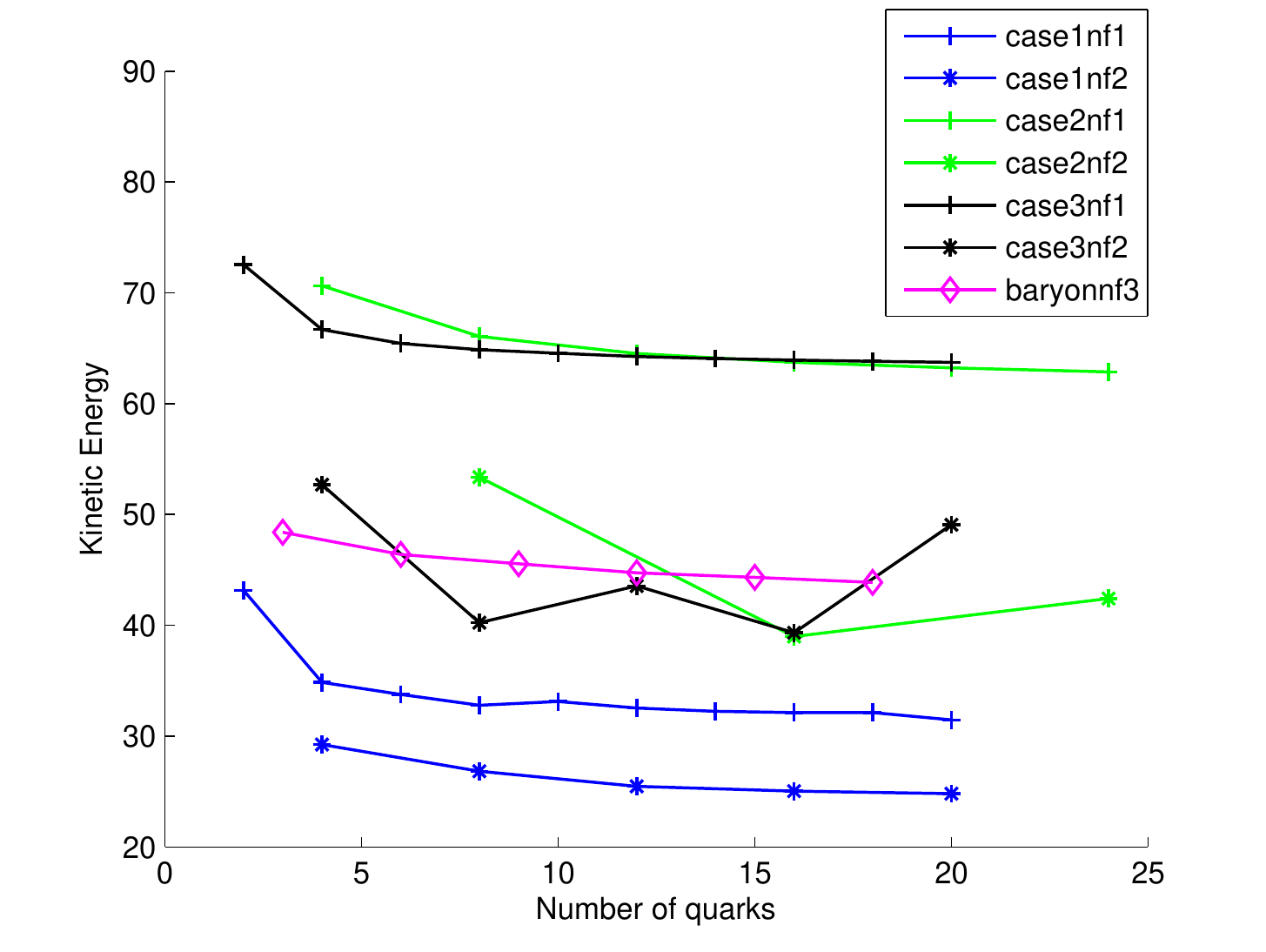}
\caption{Kinetic energy per quark (in MeV) versus quark content.}
\label{fig8}
\end{center}
\end{figure}

\begin{figure}
\begin{center}
\includegraphics[width=3.8in]{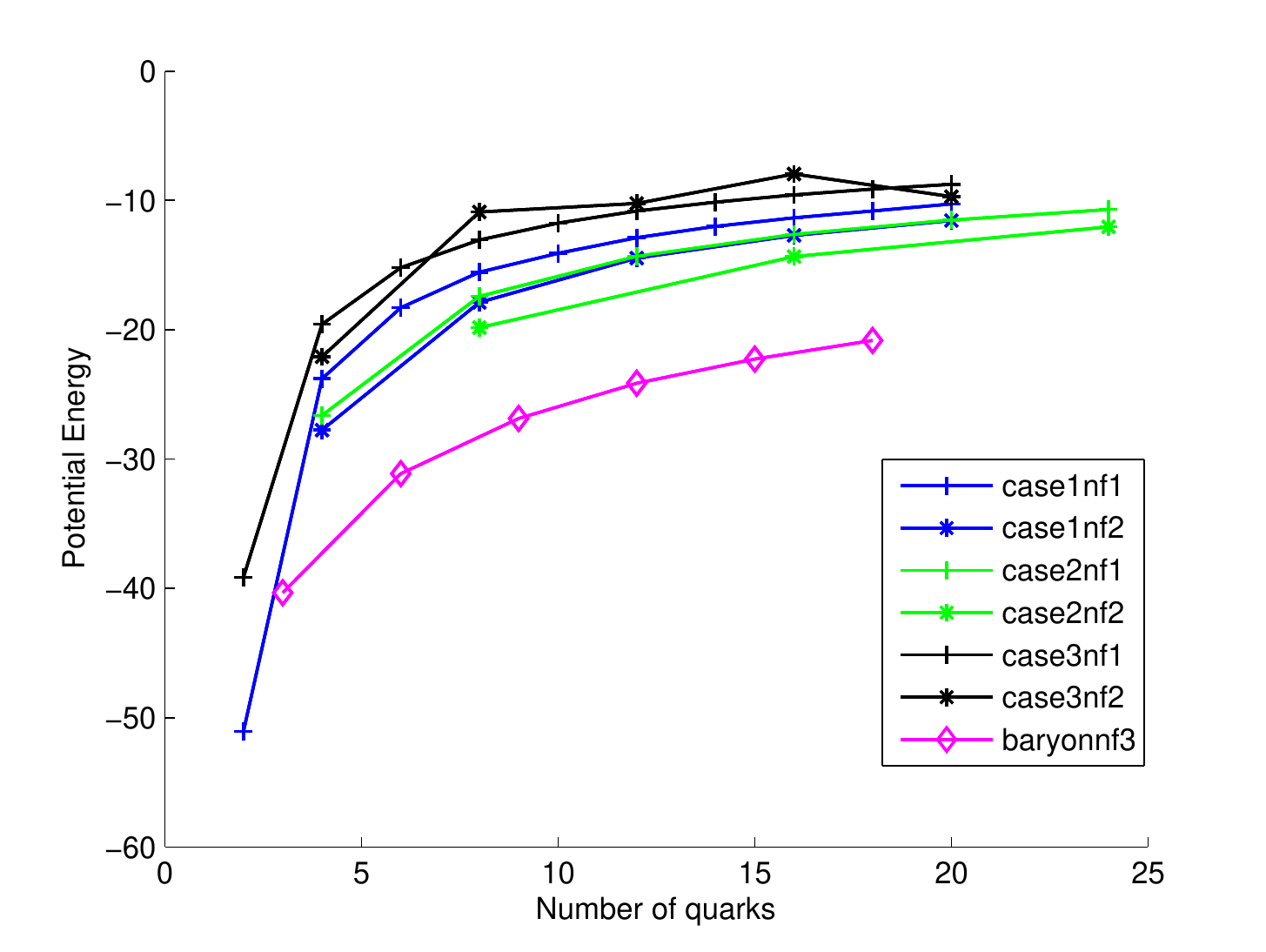}
\caption{Potential energy per quark (in MeV) versus quark content.}
\label{fig9}
\end{center}
\end{figure}

\begin{figure}
\begin{center}
\includegraphics[width=3.8in]{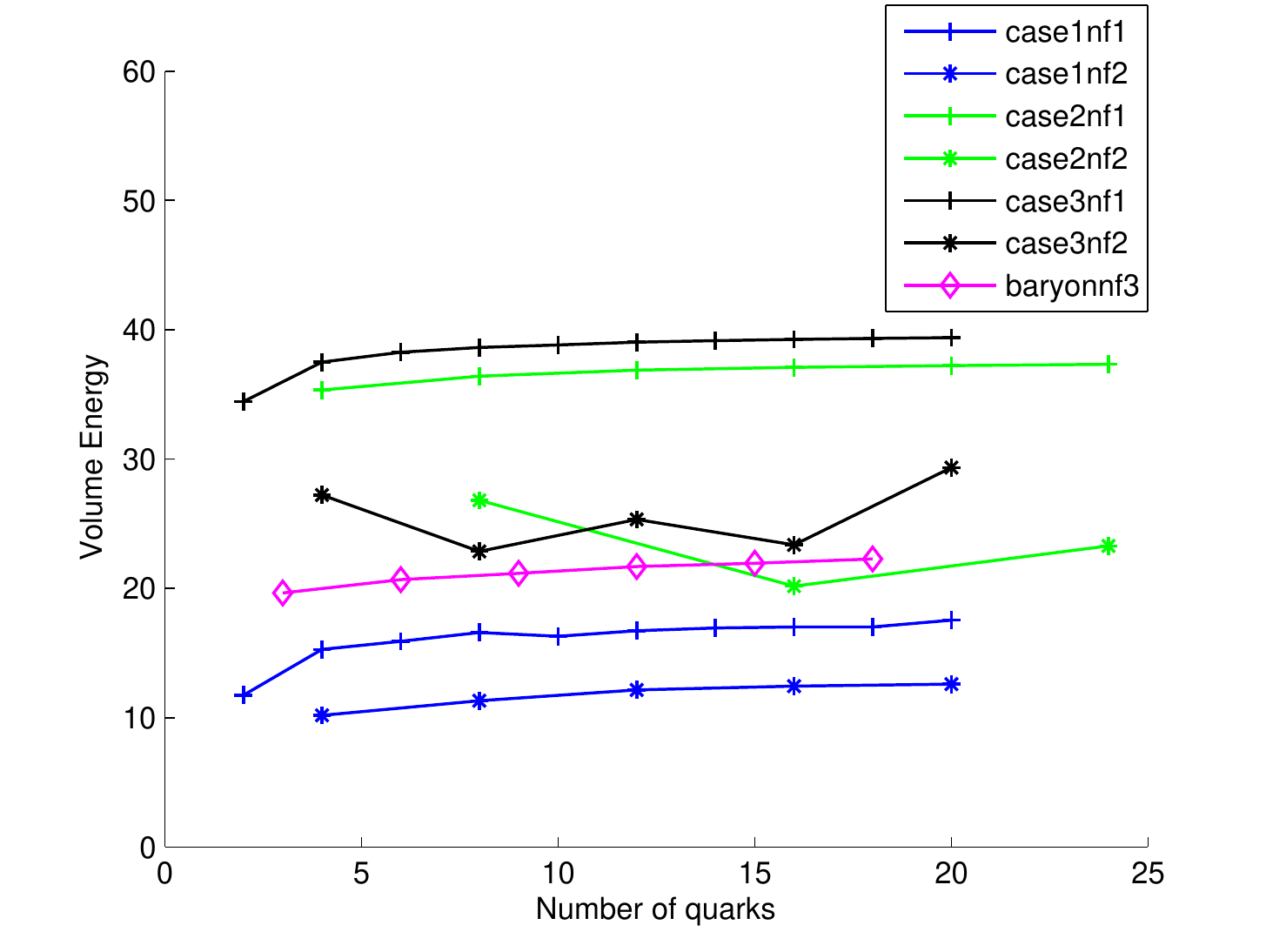}
\caption{Volume energy per quark per quark (in MeV) versus quark content.}
\label{fig10}
\end{center}
\end{figure}

\begin{figure}
\begin{center}
\includegraphics[width=3.8in]{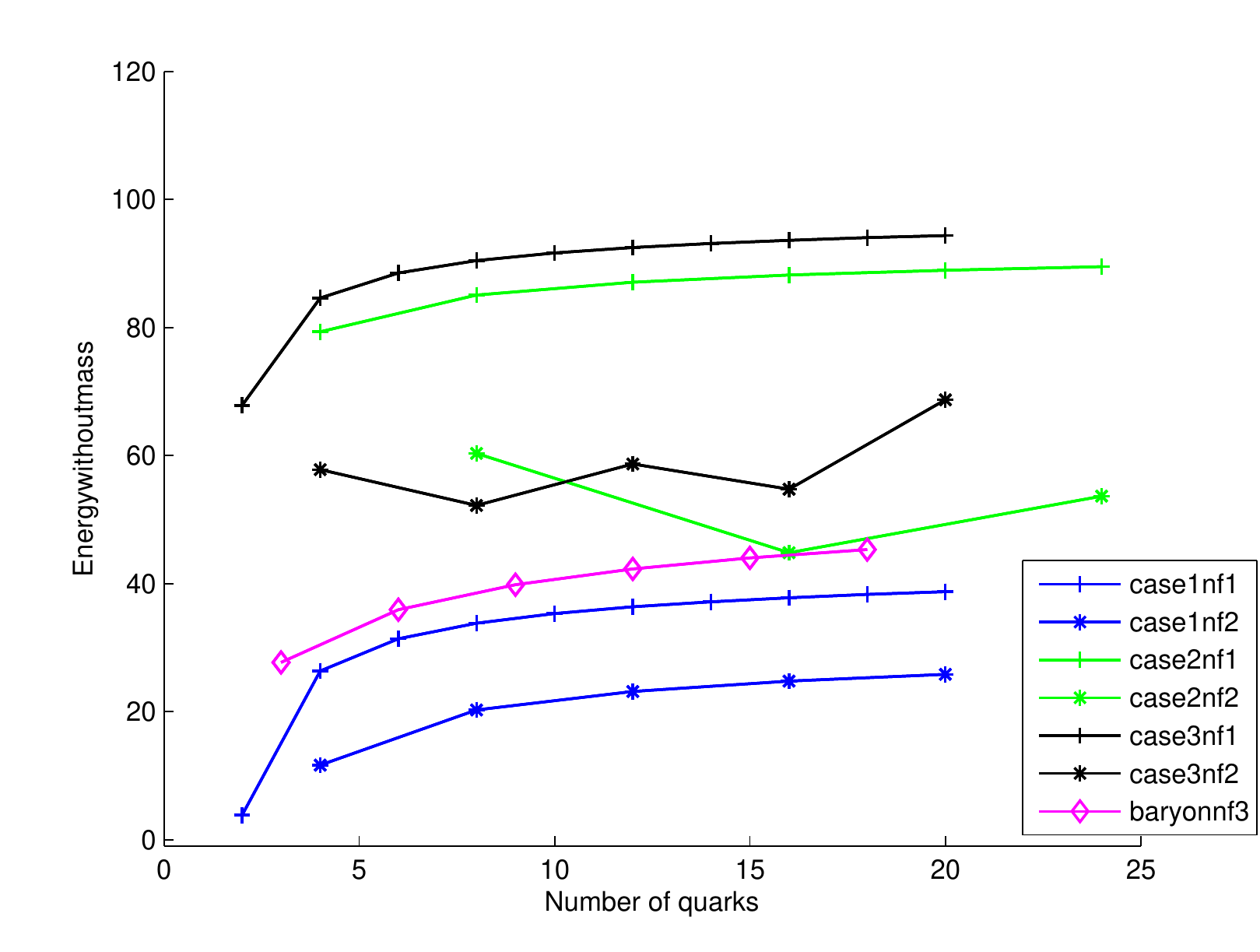}
\caption{Total energy per quark without mass terms (in MeV) versus quark content.}
\label{fig11}
\end{center}
\end{figure}

Fig.~\ref{fig7} includes 10 possibilities for the physical radius of the various mesons, compared to a generic baryon with three equal mass flavors. The radius is plotted versus quark number and compared with a generic baryon with three degenerate light flavors. We observe that the curve of the radius plot for each case tends to flatten out for larger numbers of quarks. {\bf case1nf1xmax} refers to quark families of charmonium with no degeneracy, $g=1$. In this case all the charm quarks have the same spin and hence cannot occupy the same state. {\bf case1nf2xmax} is instead the plot of the charmonium family with $g=2$. In this case, spin up and down is assigned to a pair of charm quarks. The physical radius of {\bf case1nf1xmax} being larger than {\bf case1nf2xmax} reflects this fact. Note that the dotted lines refer to the inner boundary associated with the charmed quark in Cases 2 and 3. For Case 2 the difference between dotted and continuous lines is the largest. {\bf case2nf1x1} and {\bf case2nf1x2} refer to the radius plot of the outer and inner boundary of the quark family of $Z$-mesons with $g_2=1$, respectively, while {\bf case2nf2x1} and {\bf case2nf2x2} refer to the same type of plot with $g_2=2$,. Similarly for Case 3. The $Z$ and $D$-meson family members are found to have equally large outer boundaries. 

There are 3 types of energies in this model: kinetic, potential and volume. The kinetic energy per quark depends strongly on the meson family, as seen in Fig.~\ref{fig8}. Six separate lines are given (three meson cases and two degeneracies) and compared with the generic baryon. We see that the energy per quark is relatively small and tends to decrease slowly, which seems to provide some justification for this nonrelativistic model. Fig.~\ref{fig9} shows the corresponding graph for the potential energy. These energies curve upward and level off as more quark pairs are introduced. The volume energies in Fig.~\ref{fig10} are relatively flat. The {\bf case2nf2} and {\bf case3nf2} results in these figures deserve some extra comments. These lines are determined by the most degenerate state, and thus the smallest energy per quark, available for the given quark content. If we denote the total number of quarks and antiquarks as $N$ ($=2\eta$), the $N=$ 8, 16 and 24 quark cases for {\bf case2nf2}, which all have $g_2=2$, are associated with $g_1=2$, $g_1=4$ and $g_1=3$, respectively. Likewise, the $N=$ 4, 8, 12, 16 and 20 quark cases for {\bf case3nf2} are associated with $g_1=2$, $g_1=4$, $g_1=3$, $g_1=4$ and $g_1=2$, respectively.

Fig.~\ref{fig11} is our final result. It shows the total energy per quark without the mass term, i.e., the sum of kinetic, potential and volume energy, plotted against the quark content. The generic baryon rises slowly for increasing quark content, implying these are unstable; i.e., a higher quark content state can decay into lower members of the same family. The Case 1 mesons rise quickly and then continue the rise more slowly; these are also unstable. Similarly, the Case 2 and 3 nondegenerate mesons show increases in energy per quark similar to Case 1. In contrast, for both the Case 2 and 3 degenerate families, there are initial downward, then upward tendencies. In fact, the downward jumps from $N=4$ to $N=8$ to $N=16$ for Case 2 implies octaquark and hexadecaquark states which should be stable against decay into lower family members. In addition, the downward jumps from $N=2$ to $N=4$ to $N=8$ for Case 3 implies stable tetraquark and octaquark states.

Our Case 3 tetraquark results can be compared to a lattice calculation. Ref.~\citenum{Nilmani} has preliminary two-point function results for quark flavor content $\bar{c}\bar{c}ud$. Although their calculations are limited to large dynamical pion masses $\sim550$ MeV, their results show binding of perhaps 15 MeV. This would be expected to become more strongly bound at physical quark mass. Our Case 3 results from Fig.~\ref{fig11} show a downward jump in energy per quark from $N=2$ to $N=4$ of about 10 MeV, corresponding to a total binding energy of about 40 MeV. The comparison is very encouraging.

\section{Conclusions and Acknowledgements}\label{sec9}

We have motivated a description of multi quark-pair meson states using the Thomas-Fermi statistical approach. After specifying the explicit interactions and summing on colors, we have formulated system interactions and energies. We have investigated three cases of mesonic states: charmonium family (Case 1), $Z$-meson family (Case 2) and $D$-meson family (Case 3). We have not yet included explicit spin interactions in our model, but we can take one level of degeneracy into account in our two-TF function construction. As we have said before, the goal of such a program is to prepare the way for more detailed lattice calculations.

In this study, we have surveyed the physically relevant parameter space of the TF quark model, looking for relative stability and connections to known phenomenology. We have observed interesting patterns of single quark energies. Similar to our findings for baryons, the energy per quark is slowly rising for the Case 1 mesons, implying family instability. Our Case 2 and 3 findings are the most interesting. In these cases we see an actual {\it decrease} in the energy of introduced quark pairs. For Case 2 this would superficially indicate that stable octaquark and hexadecaquark versions of the $Z$-meson exist, and that stable tetraquark and octaquark versions of the charmed $D$-meson exist. The preliminary results of Ref.~\citenum{Nilmani} gives evidence from the lattice that Case 3 meson tetraquarks are indeed stable against decay into $D\,D^*$ mesons.

Our first order of business as we extend the model will be the inclusion of $b$-meson states. In addition, we also need to evaluate and include explicit quark spin interactions to bring our meson model up to the same level of development as the TF baryon model. These interactions can be determined from the nonrelativistic ground state wave functions of these states and the associated TF function probability. Further extensions of this model would be to examine systems with heavy central charge, similar to atomic systems, or baryonic states such as generalized pentaquark families.

We thank the Baylor Quantum Optics Initiative and the University Research Committee of Baylor University for their partial support of this project. We thank the organizers of the {\it Schwinger Centennial Conference} for their invitation. We also thank N.~Mathur for helpful discussions.

\end{document}